\documentclass[useAMS,usenatbib]{mn2e}
\usepackage{graphicx}
\usepackage{times}

\title[LMT observations of cluster-lensed SMGs]{Early Science with the Large Millimeter Telescope: 
Observations of dust continuum and CO emission lines of cluster-lensed submillimetre galaxies at $\bmath{z=2.0-4.7}$}
\author[J. A. Zavala et al.]{J. A. Zavala\thanks{E-mail: zavala@inaoep.mx}$^{1}$, M. S. Yun$^{2}$, I. Aretxaga$^{1}$, D. H. Hughes$^{1}$, G. W. Wilson$^{2}$, J. E. Geach$^{3}$,
\newauthor E. Egami$^{4}$,  M. A. Gurwell$^{5}$, D. J. Wilner$^{5}$, Ian Smail$^{6}$, A. W. Blain$^{7}$, S. C. Chapman$^{8}$,  
\newauthor K. E. K. Coppin$^{3}$, M. Dessauges-Zavadsky$^{9}$, A. C. Edge$^{10}$,   A. Monta\~na$^{11,1}$,  K. Nakajima$^{9}$,  
\newauthor  T. D. Rawle$^{12}$, D. S\'anchez-Arg\"uelles$^{1}$, A. M. Swinbank$^{10}$,  T. M. A. Webb$^{13}$,   M. Zeballos$^{1}$\\
$^{1}$Instituto Nacional de Astrof\'{i}sica, \'{O}ptica y Electr\'{o}nica (INAOE),
Luis Enrique Erro 1, Sta. Ma. Tonantzintla, Puebla, Mexico\\
$^{2}$Department of Astronomy, University of Massachusetts, Amherst, MA 01003, USA\\
$^{3}$Center for Astrophysics Research, Science and Technology Research Institute,
University of Hertfordshire, Hatfield, AL10 9AB, UK\\
$^{4}$Steward Observatory, University of Arizona, 933 North Cherry Avenue, Tucson, AZ 85721, USA\\
$^{5}$Harvard-Smithsonian Center for Astrophysics, 60 Garden street, Cambridge, MA 02478, USA\\
$^{6}$Centre for Extragalactic Astronomy, Department of Physics, Durham University, South Road, Durham DH1 3LE, UK\\
$^{7}$Department of Physics and Astronomy, University of Leicester, University Road, Leicester LE1 7RH, UK\\
$^{8}$Department of Physics and Atmospheric Science, Dalhousie University, 6310 Coburg Rd., Halifax, NS B3H 4R2, Canada\\
$^{9}$Observatoire de Gen\`{e}ve, Universit\'{e} de Gen\`{e}ve, 51 Ch. des Maillettes, 1290 Versoix, Switzerland\\
$^{10}$Institute for Computational Cosmology, Department of Physics, Durham University, South Road, Durham DH1 3LE\\
$^{11}$Consejo Nacional de Ciencia y Tecnolog\'ia (CONACyT), Av. Insurgentes Sur 1582, 03940, D.F., Mexico \\
$^{12}$ESAC, ESA, PO Box 78, Villanueva de la Canada, E-28691 Madrid, Spain\\
$^{13}$Department of Physics, Ernest Rutherford Building, 3600 rue University, McGill University, Montreal, QC H3A 2T8, Canada\\}

\begin{document}

\date{Accepted 2015 June 15. Received 2015 June 15; in original form 2014 December 23}
%

\maketitle

\label{firstpage}

\begin{abstract}
We present Early Science observations with the Large Millimeter Telescope, 
AzTEC 1.1 mm continuum images and  wide bandwidth spectra ($73-111$ GHz)
acquired with the Redshift Search Receiver, towards four bright lensed submillimetre
galaxies identified through the {\it Herschel} Lensing Survey-snapshot and the 
Submillimetre Common-User Bolometer Array-2 Cluster Snapshot Survey. This pilot
project studies the star formation history and the physical properties of the 
molecular gas and dust content of the highest redshift galaxies identified 
through the benefits of gravitational magnification. We robustly detect dust 
continuum emission for the full sample and CO emission lines for three of the targets. We
find that one source shows spectroscopic multiplicity  and is a blend of three 
galaxies at different redshifts ($z=2.040$, 3.252 and 4.680), reminiscent of previous
high-resolution imaging follow-up of unlensed submillimetre galaxies, but with a completely different
search method, that confirm recent theoretical predictions of physically unassociated 
blended galaxies. Identifying  the detected lines as 
$^{12}$CO (J$_{up}=2-5$) we derive spectroscopic redshifts, molecular gas masses, and dust 
masses from the continuum emission. The mean H$_2$ gas mass of the full sample is 
$(2.0 \pm 0.2)\times10^{11} M_{\sun}/\mu$, and the  mean dust mass is 
$(2.0\pm0.2)\times10^{9} M_{\sun}/\mu$, where $\mu\approx2-5$ is the expected lens amplification.
Using these independent estimations we infer a gas-to-dust ratio of $\delta_{\rm GDR}\approx55-75$,
in  agreement with other measurements of submillimetre galaxies.  Our magnified high-luminosity
galaxies fall on the same locus as other high-redshift submillimetre galaxies,  extending the  
$L'_{\rm CO}$-$L_{\rm FIR}$ correlation observed for local luminous and ultraluminous infrared
galaxies to higher far-infrared and CO luminosities.
\end{abstract}


\section{Introduction}
Observations of nearby galaxies with the {\it Infrared Astronomical Satellite (IRAS)},
along with observations of the far-IR (FIR)/submillimetre background with the
{\it Cosmic Background Explorer (COBE)}, 
showed that the Universe
emits a comparable energy density at infrared (IR) and submillimetre wavelengths
as it does at optical and ultraviolet wavebands (e.g. 
\citealt{1987ARA&A..25..187S}; \citealt{1996A&A...308L...5P};
\citealt{1998ApJ...508..123F}). A breakthrough in resolving this 
background came with the discovery of a large population of bright sources at
high redshift through single-dish telescopes  observations at submillimetre 
wavelengths (e.g. \citealt{1997ApJ...490L...5S}; \citealt{1998Natur.394..248B}; \citealt{1998Natur.394..241H}). 
These (sub)millimetre-selected galaxies (hereafter SMGs) are characterized by large
FIR luminosities ($\ga10^{12}$ L$_{\sun}$), large
star formation rates (SFRs, $\ga300$ M$_{\sun}$ yr$^{-1}$), large gas
reservoirs ($\ga10^{10}$ M$_{\sun}$), and a number density that is high
compared to local ultra-luminous IR galaxies (\citealt{2005ApJ...622..772C}; see also review by
\citealt{2014arXiv1402.1456C}).

However, identifying and understanding the nature of these  sources
has proven to be challenging due to the low angular resolution of 
single-dish telescopes and the faintness of these galaxies in the rest-frame
optical and ultraviolet bands, which makes  the association with the correct 
counterpart difficult  (\citealt{2002PhR...369..111B}). High-resolution radio/mid-infrared
observations have often been used to identify  counterpart galaxies 
on $1-2$ arcsec scales (e.g. \citealt{2002MNRAS.337....1I}; \citealt{2008ApJ...675.1171P}), providing the precision 
needed to target optical redshifts, which are essential to understand the 
physical nature of these sources. However, recent submillimetre interferometric observations 
have shown that resourcing to surrogate wavelengths could miss $\sim45$ per cent of SMGs, and of those which are claimed 
to be identified, approximately one third are incorrect (\citealt{2012A&A...548A...4S}; \citealt{2013ApJ...768...91H};
\citealt{2015ApJSimpson}). Furthermore,
the use of these wavelengths suffers from a well-known systematic bias against 
high-redshift ($z \ga 3$ ) sources and therefore our understanding of these  galaxies could be incomplete. 

Millimetric spectroscopy provides the best way to determine redshifts
for this population of galaxies, since  they are 
expected to have luminous emission lines in the millimetre bands. Furthermore, 
the cold molecular phase of the interstellar medium (ISM) is an excellent 
tracer of the molecular gas reservoir, and the overall galaxy dynamics. For instance,
CO(1--0) has been extensively used to map the molecular gas content of this kind of galaxies 
(e.g. \citealt{2010ApJ...714.1407C}; \citealt{2010ApJ...723.1139H}; \citealt{2010MNRAS.404..198I}; 
\citealt{2011ApJ...739L..31R}; \citealt{2013MNRAS.429.3047B}; see also review by  \citealt{2013ARA&A..51..105C}).
When combined with the SFRs (or FIR luminosities),
gas masses also provide estimations of the star formation efficiencies (SFE), which is an
important parameter for models of galaxy evolution (e.g. \citealt{2014A&A...562A..30S};
\citealt{2014arXiv1408.0816D}).

However, CO spectroscopy is very time consuming, when retuning receivers to search 
for molecular line emission,  requiring entire 
nights to detect a single source, and in some cases with no successful detection
(e.g. \citealt{2005MNRAS.359.1165G}; \citealt{2014MNRAS.443L..54H}). This method has 
only recently become competitive  with the increased bandwidth of (sub)mm facilities,
but is still  not yet sufficiently efficient to obtain redshifts for substantially 
large samples of unlensed SMGs (\citealt{2013MNRAS.429.3047B}). On the other hand, 
CO line observations for strongly lensed systems can be obtained more easily, 
representing one route to study the molecular gas content of SMGs 
(e.g.  \citealt{2009A&A...496...45K}; \citealt{2011ApJ...740...63C}; \citealt{2012ApJ...752..152H};
\citealt{2013ApJ...767...88W}). Although in some cases, differential magnification 
could affect the properties derived from observations (e.g. line ratios, \citealt{2012MNRAS.424.2429S}),
cluster lensing provide us more reliable results since is rare to find this kind of magnification gradients.

Here we present new observations of a sample of four cluster-lensed SMGs,
discovered by the {\it Herschel} Lensing Survey-snapshot (HLS-snapshot, Egami et al. in preparation;
see also \citealt{2010A&A...518L..12E}) 
and subsequently detected with the Submillimetre Common-User Bolometer Array-2 
(SCUBA-2, \citealt{2013MNRAS.430.2513H}) on the James Clerk Maxwell Telescope (JCMT)
as part of the SCUBA-2 Cluster Snapshot Survey 
(S2CSS, Geach et al. in preparation). The new observations comprise imaging from the 
1.1 mm continuum camera AzTEC (\citealt{2008MNRAS.386..807W}) and  the wide
bandwidth ($73-111$ GHz) spectrometer Redshift Search Receiver (RSR, 
\citealt{2007ASPC..375...71E}) installed  on the Large Millimeter Telescope 
{\it Alfonso Serrano} (LMT\footnote{www.lmtgtm.org}, \citealt{2010SPIE.7733E..12H}),
located on the summit of Volc\'an Sierra Negra (Tlilt\'epetl), Mexico,
at an altitude of $\sim4600$ m. Combining the  high sensitivity and bandwidth of the LMT 
instruments with the higher flux density of amplified galaxies, the detection of CO 
molecular lines and dust continuum can be reached in reasonable times, allowing 
us to  derive spectroscopic redshifts, and therefore to inquire into the dust and gas properties
in these galaxies.

In \S2, we describe the 
sample selection, observations and data reduction. In \S3, we describe the line 
identification, redshift estimations, and gas and dust properties. In \S4,
we discuss the implication of these results for the gas-to-dust ratio and SFE. Our results
are summarized in \S5.

All calculations assume a $\Lambda$ cold dark matter cosmology with
$\Omega_\Lambda=0.68$, $\Omega_m=0.32$, and
$H_0=67$ kms$^{-1}$Mpc$^{-1}$ (\citealt{2013arXiv1303.5076P}).

\section{OBSERVATIONS AND DATA REDUCTION}
\subsection{Sample selection and SCUBA-2/JCMT data}\label{sample_selection}
The four bright SMGs targeted here were originally discovered by the HLS-snapshot survey,
which has obtained shallow but nearly confusion-limited Spectral and Photometric Imaging Receiver (SPIRE)
images of 279 Massive Cluster Survey (MACS) clusters. Three of the four galaxies have a SPIRE flux densities of more
than 100 mJy at either 250, 350 or 500 $\mu$m, and have subsequently produced bright detections in the S2CSS.
The remaining target (HLS J102225.9+500536) was detected by HLS below the 100 mJy threshold but was comparably bright at 850 $\mu$m.
The analysis of the  {\it Herschel} data and their optical-IR counterparts 
will be presented elsewhere by the HLS team.

\begin{figure*}
\includegraphics[width=170mm]{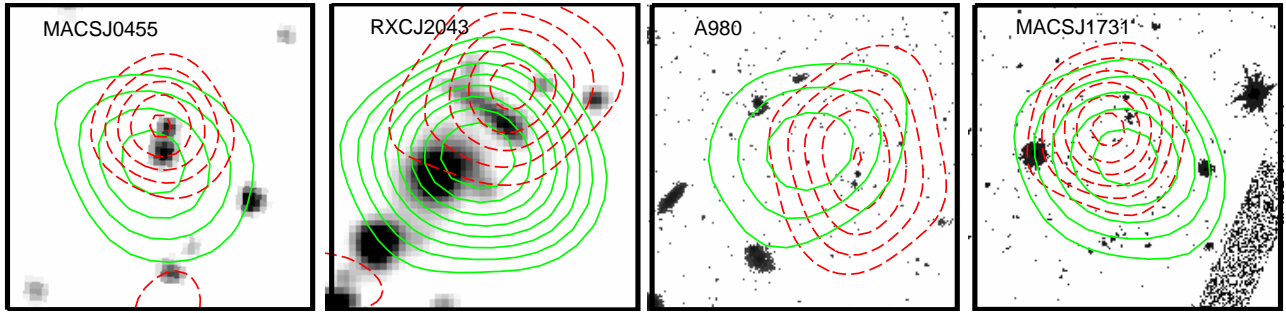}
\caption{From left to right, $30\times30$ arcsec$^2$ Infrared Array Camera (IRAC)/{\it Spitzer} 
images in the 3.6 $\mu$m band at the SCUBA-2 position of  
HLS J045518.0+070103 and HLS J204314.2--214439 
and {\it HST} images in the  $F606W$ band at the position of HLS J102225.9+500536
and HLS J173140.8+225040. The SCUBA-2 850 $\mu$m and
AzTEC 1.1mm S/N contours are also shown in green (solid line)
and red (dashed line), respectively. Contour levels start at 
1.5$\sigma$ and are spaced in steps of 1.5$\sigma$. See the electronic
edition of the journal for a colour version of this figure.}
\label{spitzer}
\end{figure*}

\begin{table*}
\caption{LMT observing log for data taken during the Early Science Phase in 2014. Column 1, name of
the targeted galaxies; columns 2, 3, and 4: dates, opacities and integration times of AzTEC 
observations; columns 5, 6, and 7: dates, opacities and integration times 
of RSR observations.}
\begin{tabular}{llccclcc}
\hline
 &\multicolumn{3}{c}{AzTEC}&&\multicolumn{3}{c}{RSR}\\
ID &Date&Opacities&$t_{int}$&&Date&Opacities&$t_{int}$\\
   &    &   ($\tau_{225GHz}$) & (min) &  &  &($\tau_{225GHz}$) & (min)\\
\hline
HLS J045518.0+070103&Feb 10&0.12&20& &Feb 19, Mar 02, Apr 02 &0.06, 0.17, 0.26--0.29&45, 45, 20\\

HLS J102225.9+500536&Jan 31&0.10&20& &Mar 27, 29, Apr 02, 07 &0.17, 0.15, 0.20, 0.15&10, 60, 50, 35\\
HLS J173140.8+225040&Mar 21, May 18&0.12, 0.18&4, 10& &Apr 21, May 03&0.30, 0.30&30,75\\
HLS J204314.2--214439&May 14, 16&0.20, 0.22&10, 10 & &May 08, 12, 18, 25, 26 &0.46-0-57, 0.29-0.32,&90, 30, 50, 60, 30 \\
             &            &          &      & &        & 0.13-0.15, 0.42--0.44, 0.35 & \\

\hline
\label{lmt_observatios}
\end{tabular}
\end{table*}

S2CSS exploited poor weather conditions at the JCMT (opacities of
$\tau_{\rm 225GHz}=0.12$-$0.2$) to search for  bright ($S_{850\mu m}>30$ mJy) gravitationally lensed SMGs, comparable
to the `Cosmic Eyelash' (\citealt{2010Natur.464..733S}), in the cores of hundreds of
rich clusters of galaxies,  primarily selected from the MACS and SMACS surveys (e.g. 
\citealt{2001ApJ...553..668E}; see also target selection of the HLS in \citealt{2010A&A...518L..12E}).
The `Cosmic Eyelash' is a typical SMG at $z\approx2.3$ with an intrinsic flux of $S_{870}\sim3$ mJy, which is magnified 30 times by a 
foreground cluster, appearing as an $S_{870}\sim106$ mJy source.  The high magnification and spatially-resolved structure allows a very detailed
study of the internal structure and chemistry of this galaxy on $<100$ pc scales.
Although rare, the bright nature of these sources makes them detectable at high 
significance even when observing conditions are relatively poor. For each  cluster target
a 30\,minute DAISY scan map was obtained with SCUBA-2, which allowed the detection of luminous SMGs within
a few arc minutes from the cluster cores (depending on source brightness). The data were reduced 
with the Sub-Millimetre User Reduction Facility ({\sc smurf}; \citealt{2013MNRAS.430.2545C}), following a 
procedure described in \citet{2013MNRAS.432...53G} where the map making was optimized for point-source
detection. Typical root mean squared noise (rms) noise in individual 850-$\mu$m maps was 5--10 mJy (see Table \ref{fotometria}). 

We conducted observations  with AzTEC and RSR on the LMT towards a sample of four 
850 $\mu$m sources that represent some of the most secure ($>5\sigma$) detections
within S2CSS around the clusters MACS J0455.2+0657, Abell 980 (MACS J1022.4+5006), MACS J1731.6+2252, 
and RXCJ2043.2--2144 (MACS J2043.2--2144) at $z=0.425$, 0.158, 0.366, and 0.204, respectively
(\citealt{1998MNRAS.301..881E}, \citeyear{2010MNRAS.407...83E}; \citealt{2004A&A...425..367B}; \citealt{2012MNRAS.420.2120M}).
Postage stamps of {\it Spitzer} or {\it Hubble Space Telescope (HST)} images, and overlaid with SCUBA-2 and AzTEC contours at these positions
are shown in Fig. \ref{spitzer}. Details of these bright galaxies, HLS J045518.0+070103,
HLS J102225.9+500536, HLS J173140.8+225040, and HLS J204314.2--214439, and their fluxes are given in Table \ref{fotometria}.

\subsection{AzTEC/LMT}

Observations  were obtained using the 1.1 mm continuum camera AzTEC on the LMT between 2014 January and May (see 
Table \ref{lmt_observatios}). During this Early Science Phase operation, only the inner
32-m diameter section of the telescope surface is illuminated, leading to an effective
beam size of $\sim8.5$ arcsec, a factor of $\sim2$ better than the SCUBA-2 850 $\mu$m-beam.
The scanning technique provided a Lissajous pattern covering a 1.5 arcmin diameter region 
with uniform noise in the map centred at the SCUBA-2 position. Observations were conducted over
several observing nights with an opacity range of $\tau_{\rm 225GHz}= 0.07-0.18$ and a 
typical integration time of $\sim20$ min per source, achieving an rms of 2--3 mJy  (see Table \ref{lmt_observatios})
in the individual maps.

Each individual AzTEC observation is composed of a set of timestreams which store all
the bolometer sky signals, the projected position of the array on the sky and the 
weather conditions. Calibration is performed using observations of the asteroids Ceres and Pallas.  
To generate the final maps from the raw data we use the AzTEC 
Standard Pipeline described in detail by \citet{2008MNRAS.385.2225S}. 

\begin{figure*}
\includegraphics[width=170mm]{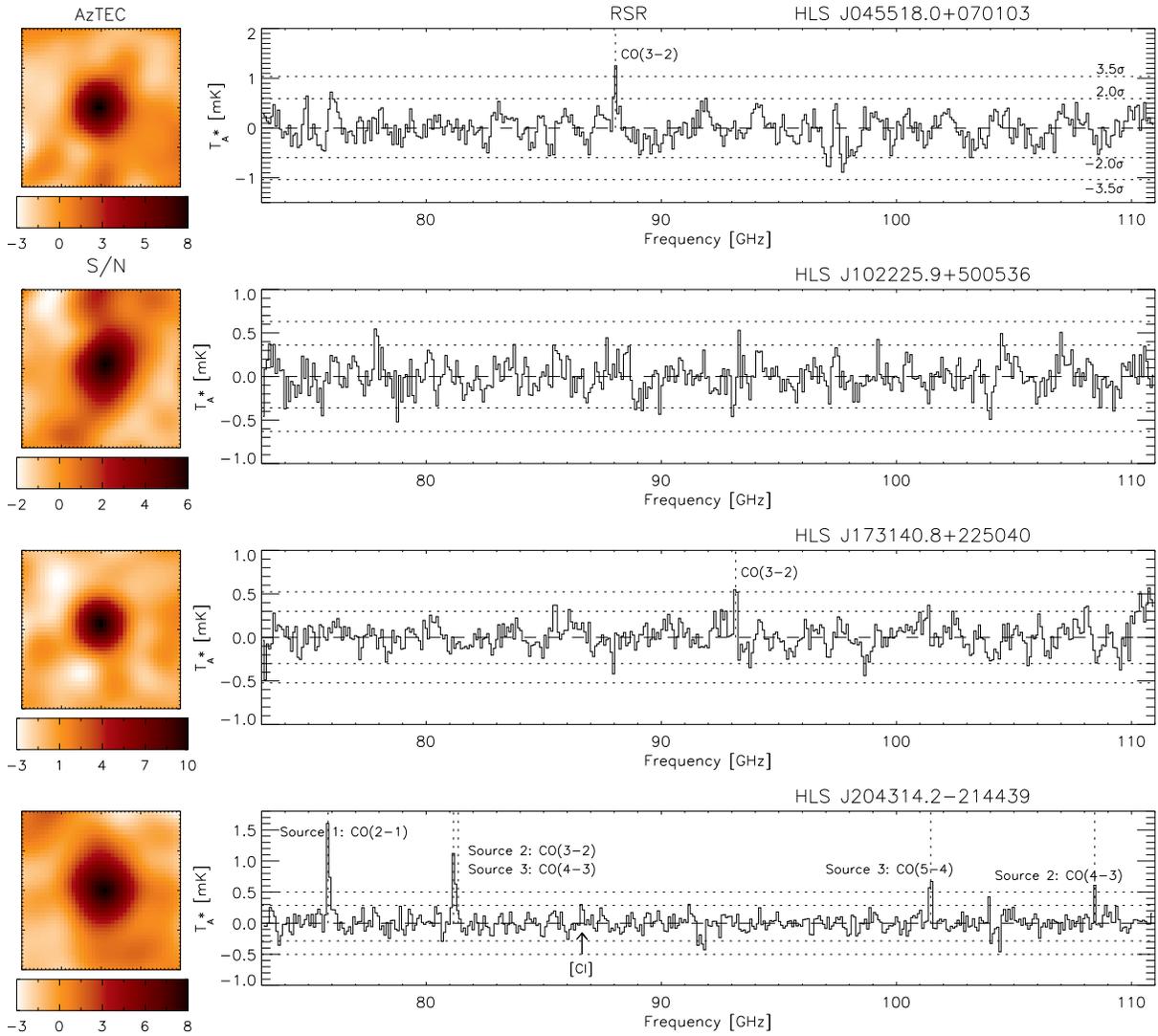}
\caption{Left: $40\times40$ arcsec$^2$ AzTEC/LMT 1.1 mm S/N maps. 
All the  maps are centred at the AzTEC positions listed in Table \ref{fotometria},
and have been scaled for display purposes. Right:  RSR/LMT spectra of the four HLS/S2CSS targets
observed with the LMT. The RSR/LMT spectra have been rebinned into 3 pixels bins for  better visualization. 
We mark with horizontal lines the 2 and 3.5$\sigma$ average noise level for each spectrum, where 1$\sigma$ has 
been calculated as the standard deviation of the whole binned spectrum after removing the identified lines.
The transitions detected above the adopted threshold detection are marked with vertical dashed lines. In
the spectrum of HLS J204314.2-214439 we have identified contributions from three different galaxies at 
different redshifts. We also mark with an  arrow the expected position for the [CI] transition at $z=4.680$
(see \S\ref{MACSJ2043}). See the electronic edition of the journal for a colour version of this figure.}
\label{imagenes}
\end{figure*}

In Table \ref{fotometria} we summarize the 1.1 mm photometry measurements from the AzTEC maps shown
in  Fig. \ref{imagenes}. The AzTEC positions, which, for these observations, have a typical uncertainty of 
$\sim2$ arcsec, are consistent with the SCUBA-2 positions 
within the 95 per cent confidence intervals, except for the case of HLS J204314.2--214439. However, as discussed in \S 
\ref{MACSJ2043}, our RSR spectrum has revealed three  sources in the beam for this target and, therefore, the
estimated positional uncertainty could be underestimated.

\begin{table*}
\caption{Summary of SCUBA-2 and AzTEC photometry. Column 1: name of the targeted galaxies; columns 2, 3, 4,
and 5: position, 95 per cent confidence interval for positional uncertainties following the description 
by \citet{2007MNRAS.380..199I} for single sources, signal-to-noise ratio, and 850 $\mu$m flux density for SCUBA-2
observations; columns 6, 7, 8, and 9: same information for 1.1 mm AzTEC observations.}
\begin{tabular}{lcccccccc}
\hline
ID& SCUBA-2 position & $\Delta(\alpha,\delta)$  & S/N & S$_{850\mu m}$ &AzTEC position& $\Delta(\alpha,\delta)$ &  S/N & S$_{1.1mm}$ \\
& (J2000)  & (arcsec) && (mJy) & (J2000) & (arcsec)& & (mJy) \\
\hline
HLS J045518.0+070103 & $04$:$55$:$18.1$, $+07$:$01$:$01$ &3.2 &7.0   &$37\pm5$ & $04$:$55$:$18.1$, $+07$:$01$:$05$& 1.6 & 8.0 &$23\pm 3 $ \\
HLS J102225.9+500536      & $10$:$22$:$26.6$, $+50$:$05$:$39$ &4.1 &5.5   &$45\pm8$ & $10$:$22$:$26.2$, $+50$:$05$:$41$& 2.2 & 5.9 &$11\pm 2 $ \\
HLS J173140.8+225040 & $17$:$31$:$40.5$, $+22$:$50$:$35$ &2.6 &8.5   &$49\pm6$ & $17$:$31$:$40.6$, $+22$:$50$:$37$& 1.3 & 9.8 &$25\pm 3 $ \\
HLS J204314.2--214439  & $20$:$43$:$14.5$, $-21$:$44$:$37$ &1.7 &13.0  &$66\pm5$ & $20$:$43$:$14.2$, $-21$:$44$:$43$& 1.5 & 8.4 &$29 \pm 3$ \\
\hline
\label{fotometria}
\end{tabular}
\end{table*}

\subsection{Redshift Search Receiver/LMT}
Observations with RSR were  obtained between 2014 February and May. The RSR has
4 pixels arranged in a dual beam, dual polarization configuration. The four broad-band 
receivers cover instantaneously the frequency range $73-111$ GHz with a spectral resolution
of 31 MHz ($\sim100$ km s$^{-1}$ at 93 GHz). The current 32-m telescope gives us an 
effective beam size of $20$ arcsec at 110 GHz and $28$ arcsec at 75 GHz.

Observations were obtained at the SCUBA-2 positions of the four galaxies listed in 
Table \ref{fotometria}, over several observing nights with an opacity range of 
$\tau_{\rm 225GHz}= 0.13-0.57$ (see Table \ref{lmt_observatios}) and system temperatures of 
$T_{\rm sys}=90-140$ K with an average $T_{\rm sys}\approx100$ K. Pointing was performed every 
hour on bright millimetre sources, leading to a pointing accuracy (rms) $\sim2$ arcsec.

The data were reduced and calibrated using Data REduction and Analysis Methods
in PYthon ({\sc dreampy}), which is the RSR data reduction pipeline software written by G. Narayanan.
After flagging any data adversely affected by a hardware or software problem, a linear
baseline is removed from each spectrum. The final spectra were obtained by averaging
all scans using  $1/\sigma^2$ weights. The spectra in antenna temperature units 
($T_{\rm A}^*$) are converted to flux density units using the conversion factors 
6.4  and 7.6 Jy K$^{-1}$, for $73<\nu\le92$ GHz and $92<\nu<111$ GHz, respectively, which are
based on the calibration of Uranus and MWC349A, observed at EL $= 35\degr-75\degr$,
conducted in the Early Science Phase. Final spectra are shown in Fig. \ref{imagenes}.

\subsection{SMA observations}\label{SMA_data}
Observations on the cluster  RXCJ2043.2--2144 were obtained with the Submillimeter Array
(SMA)  as a follow-up programme of the {\it Herschel} Lensing Survey (HLS)  on 2012 May 8.  
After the LMT obtained a secure redshift for a $z=4.68$ galaxy in this field (see \S\ref{MACSJ2043}),
these data were reanalysed to search for the predicted  [CII] line emission
at $\nu=334.6$ GHz as a means of further confirmation.

Observations were made using six antennas in a compact configuration. An on-source
integration time of 3.9 h was obtained with the primary LO set to
342.046 GHz, using a single polarization SIS junction receiver with 4
GHz bandwidth per sideband.  This bandwidth covered the expected
frequency of the redshifted [CII] line for $z=4.68$ (rest frequency
1900.54 GHz shifted to 334.60 GHz).  The weather was very good, with
the atmospheric opacity at 225 GHz varying from 0.05 to 0.07 throughout
the observations. The synthesized beam size was  $2.4\times1.9$ arcsec$^2$. The
primary flux density calibrator was Neptune, which provided a flux scale
accurate to $\sim5$\%. Instrumental and atmospheric gains were calibrated
using complex gain calibrators J1924--292 (4.05 Jy), J2000--178 (0.46
Jy), and J2158--150 (1.66 Jy). The visibility data were resampled and
averaged to a velocity resolution of 20 km s$^{-1}$, and then binned further
to 100 km s$^{-1}$ resolution in the imaging routine. The rms noise obtained
in the final images was 15.2 mJy beam$^{-1}$ per 100 km s$^{-1}$ channel. 

These observations  showed the presence of three distinct sources with a clear emission 
line feature at 334.6 GHz. Averaging consistent
the spectra from these three positions produced the spectrum shown in
Fig. \ref{SMA_CII} (rms of 8.8 mJy), which we identify as the [CII] transition.
The analysis of the triple system and constraints to the lensing model of  RXCJ2043.2--2144 
(MACS J2043.2--2144) will be presented in a subsequent paper.

\section{RESULTS}

\subsection{RSR analysis: line identification and spectroscopic redshifts}\label{line_ident}
 
Since the main goal of these observations is to blindly search for molecular lines in our
spectra, with no previous knowledge on the expected observed frequencies for these lines, it is important
to define a criteria S/N threshold ($\xi_{\rm thresh}$) that ensures the reliability of 
detections and that maintains the false detection rate very low. 

\begin{figure}
\begin{center}
\includegraphics[width=60mm]{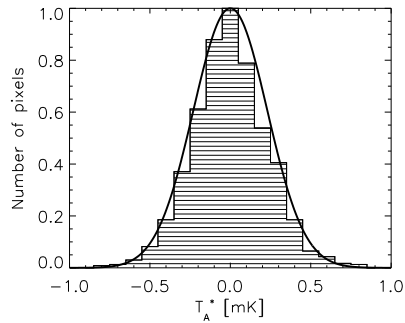}
\caption{Histogram of pixel values in the RSR/LMT spectrum of HLS J102225.9+500536, as an example of the noise properties in the spectra.
The histogram is well fitted by a Gaussian function, which is represented by the solid line. These Gaussian functions are 
used to make mock noise spectra in order to define a reliable threshold line detection.} 
\label{spectralnoise}
\end{center}
\end{figure}

To estimate the global noise of the spectra we fit the histogram of pixel values with Gaussian
distributions. As can be seen in Fig. \ref{spectralnoise}, a Gaussian fit reproduces very
well the spectrum, and can be used as an estimation of the average noise. This is true for
all the acquired spectra, achieving fits  of  $\chi_{\rm red}^2= 0.57-0.96$.
Using these Gaussian distributions, we make mock noise spectra in order to define a robust 
detection threshold. These simulations assume that the noise is uniform at all frequencies and
uncorrelated between pixels. Using 100,000 realizations we find that the probability to have 
two adjacent pixels with S/N$\ge3.5$ is less than 0.1 per cent, and therefore this S/N ratio 
could be used as a detection threshold.
To investigate the reliability of these simulations  we search for negative peaks, which 
are expected to be just noise, that satisfy this S/N threshold. We find none, either in the raw or
in binned spectra, which supports the adoption of $\xi_{\rm thresh}=3.5$ (in two adjacent pixels)
as a robust detection threshold.

 \begin{table*}
\caption{Observational parameters of the lines detected in the RSR spectra, using a blind search.
The three blended sources in the HLS J204314.2--214439 target are identified with the extensions /S1, /S2, and /S3.
Column 1: name of the source; column 2 and 3: signal-to-noise ratio; column 4: central frequency; column 5: identified transition;
column 6: redshift; column 7: integrated flux; column 8: FWHM of the line; column 9: notes.}
\begin{tabular}{lcccccccc}
\hline
Source&S/N$^a$ &S/N$^b$&$\nu_{\rm obs}$ &   Transition & $z$ & $S_{\rm CO}\Delta V$ & FWHM & Notes\\
 & &&(GHz)&   & & (Jy km s$^{-1}$) & (km s$^{-1}$)&\\
\hline
HLS J045518.0+070103     &  4.5   & 7.3    &88.051 & $^{12}$CO(3--2) & 2.927$^c$ & $4.7\pm0.8$ & $480\pm70$ & Alternative solution $z=1.618$  \\
HLS J173140.8+225040     &  4.1   & 5.7    &93.163 & $^{12}$CO(3--2) & 2.712     & $2.3\pm0.8$ & $420\pm120$& $z$ confirmed by IRAM\\
HLS J204314.2--214439/S1  &  10.3  & 15.5   &75.827 & $^{12}$CO(2--1) & 2.040     & $6.1\pm0.9$ & $530\pm60$ & $z$ confirmed by IRAM and VLT\\
HLS J204314.2--214439/S2  & 2.1$^d$& 2.0$^d$&81.328 & $^{12}$CO(3--2) & 3.252     & $0.5\pm0.3$ & $260\pm130$ & $z$ confirmed by VLT\\
                         &  5.1   & 5.7    &108.428& $^{12}$CO(4--3) & 3.252     & $1.3\pm0.4$ & $190\pm40$ &                  \\
HLS J204314.2--214439/S3  &  7.7   & 12.3   &81.178 & $^{12}$CO(4--3) & 4.680     & $4.8\pm0.6$ & $570\pm 60$& $z$ confirmed by SMA   \\
                         &  5.1   & 8.0    &101.459& $^{12}$CO(5--4) & 4.681     & $2.8\pm0.5$ & $410\pm50$ &   \\
           
\hline
\multicolumn{9}{l}{$^a$The maximum S/N of the peak pixel from either the raw or the 200 km s$^{-1}$ binned spectra. In this case, the noise has been estimated as the standard deviation}\\
\multicolumn{9}{l}{of the whole spectrum after removing the identified lines. $^b$The S/N of the integrated line in the raw spectra (100 km s$^{-1}$ pixels). $^c$Tentative solution. $^d$This line}\\
\multicolumn{9}{l}{ is not formally detected by our algorithm, but searched for after the VLT redshift solution was proposed.}\\
\label{espectros}
\end{tabular}
\end{table*}

In order to identify line candidates in our spectra, we search for peaks with S/N$\ge\xi_{\rm thresh}$,
where  the noise is estimated as the standard deviation of the whole spectrum.
If a peak is identified, we mask a region of 10 pixels centred at the position of 
the peak (since we know that the full width at half-maximum (FWHM) of lines detected in 
SMGs can be up to $\sim1000$ km s$^{-1}$, e.g. \citealt{2013MNRAS.429.3047B}). This process is
repeated until there are no more pixels with S/N$\ge\xi_{\rm thresh}$. If a line candidate is 
identified, we inspect it visually to confirm that it does not correspond to a bad baseline. In all our
spectra we do not find any such case. In order to find weaker emission lines, we rebin our spectra into 200 
km s$^{-1}$ pixels (2 pixels bins)  and repeat the process again. Once we have the position of line candidates
we carry out a Gaussian  line profile fit in order to obtain the central frequency, peak flux, and line width.

Although the general noise is well fitted by a Gaussian distribution (see Fig. \ref{spectralnoise}) 
and the baseline fluctuations in each spectrum are well confined within the  $\pm2\sigma$ noise range 
(see Fig. \ref{imagenes}), there are some baseline residuals that exceed this range or where the noise 
seems correlated. These baseline features are usually broad ($>1000$ km s$^{-1}$) and therefore could 
preclude the detection of low signal-to-noise ratio (S/N$\la5$) broad lines (albeit the fraction of SMGs with 
very broad lines, $>1000$ km s$^{-1}$,  have been found to be low, e.g. 1/32, \citealt{2013MNRAS.429.3047B}). 
Broad lines could be missed since our method is based on the detection of high S/N pixels. 
However, a different method has been developed, which exploits the full spectral information present in the
RSR data by cross-correlating the observed spectrum with a theoretical or an empirical spectral template 
(see \citealt{2007ASPC..375..174Y}). A detailed description of this method and its application to the SMG 
COSMOS AzTEC-1 is described by \citet{2015MNRASYun}, and we briefly summarize it here. A cross-correlation 
product is derived as a function of redshift from the observed spectrum  and a model spectral template. 
This model is weighted by a function which represents the relative strength of different molecular 
transitions, and for this, an empirical composite spectrum based on observed relative line strengths for 
high redshift sources (e.g., \citealt{2014ApJ...785..149S}) is adopted. This is a powerful method to derive 
redshifts which takes into account the information of many transitions which are not individually detected 
with a good S/N. We find that the redshift results are robust, and both methods give consistent results,
when the integrated S/N of the lines are sufficiently high (S/N$>5$).

\begin{figure*}
\includegraphics[width=170mm]{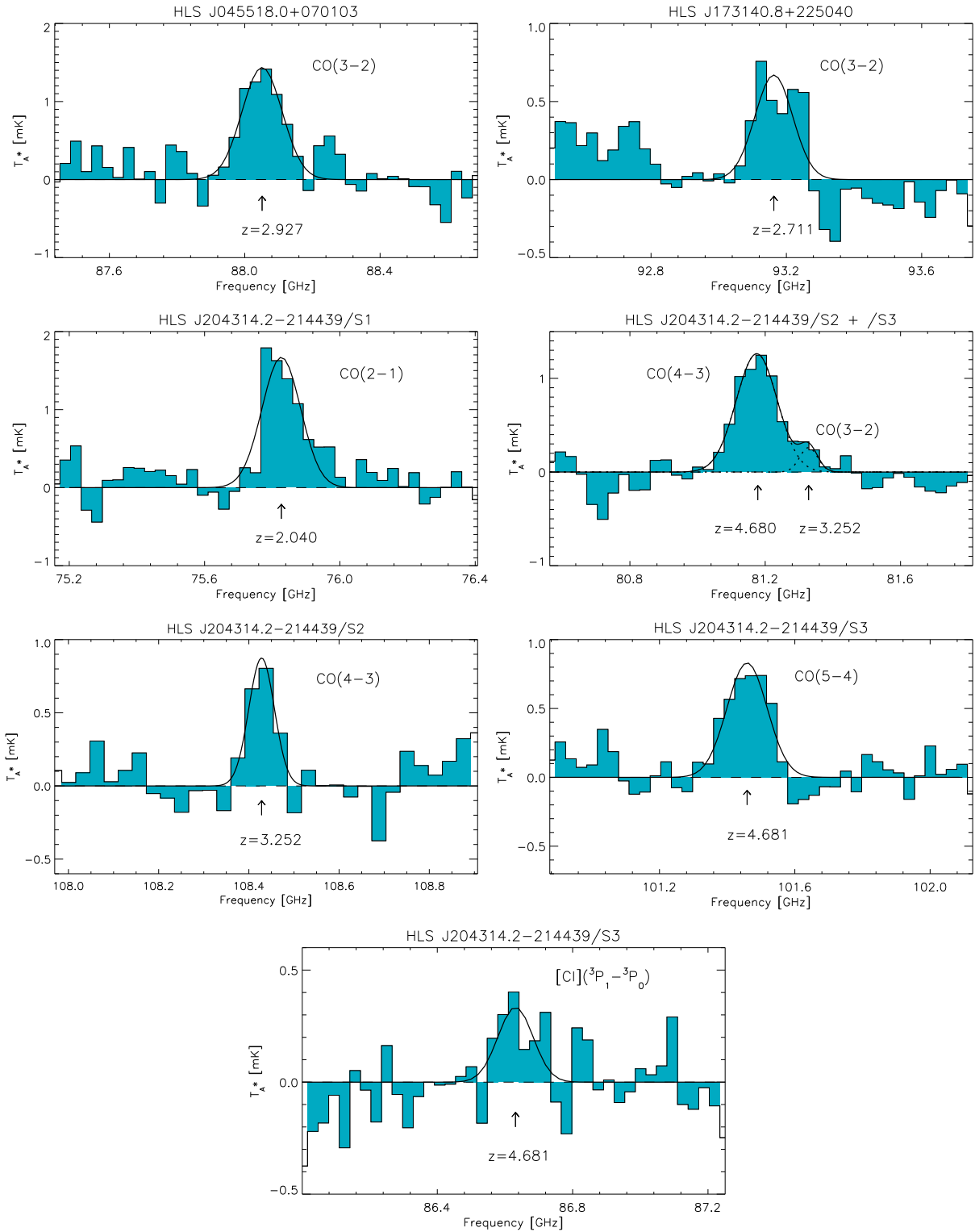}
\caption{LMT raw spectra at the position of the $^{12}$CO and [CI] lines  detected in our  RSR spectra,
best-fitting Gaussian profiles, and redshift derived from each individual line. The three blended sources in the HLS J204314.2--214439 
target are identified with the extensions /S1, /S2 and /S3. The [CI]($^3P_1 \to ^3P_0$)
transition at $z=4.680$ for HLS J204314.2--214439/S3  is only detected at $3\sigma$, and the  FWHM of the 
Gaussian fit is consistent with the line widths of the two  $^{12}$CO transitions.} 
\label{lines_all}
\end{figure*}

In Table \ref{espectros} we summarize the lines detected in our spectra and Fig. \ref{lines_all}
shows them individually. All these detections have at least one peak pixel at S/N$\ge4.0$ in the
raw  or in the binned spectra and at least one adjacent pixel with S/N$\ge3.5$ (except for a 
line that is not formally detected by our algorithm, see \S \ref{MACSJ2043}). However, a better 
estimation of the reliability of these detections is the S/N of the whole integrated line, which is also reported in 
Table \ref{espectros}. All the lines detected originally by our algorithm have integrated S/N$>5$.

In order to derive the spectroscopic redshifts we have first to associate the detected lines candidates 
with the correct molecular transitions. The simultaneous frequency coverage of the RSR  means 
at least one CO transition falls within the spectral coverage at all redshifts except for a narrow 
redshift range (the `redshift desert') in the range $0.58 < z < 1.08$. Furthermore, two or more molecular 
line transitions fall within the RSR spectral range at $z>3.15$ if we consider the possibility of a 
[CI] line detection in addition to the CO lines.
On the other hand, since we know that the spectral line 
energy distribution (SLED) peaks at J$_{\rm up}=6-5$ for typical SMGs (\citealt{2013MNRAS.429.3047B}; 
\citealt{2014ApJ...785..149S}), the absence of lines could be used to discard  
the high-redshift solutions for which we expect more than one line in our spectra. Moreover,
our selection of bright  sources through galaxy clusters, 
implies that these very bright SMGs are expected to be gravitationally lensed. 
This selection tends to exclude low-redshift  solutions, $z\la1-1.5$, due to the low 
probability of being lensed at these redshifts, as discussed by \citet{2013ApJ...767...88W}.
Taking these considerations into account, we  associate our line candidates with
different  $^{12}$CO (J$_{\rm up}= 2-5$) transitions and derive the spectroscopic redshift
for each galaxy (see Table \ref{espectros}). Below we describe briefly the  
analysis for individual targets.

\subsubsection{HLS J045518.0+070103}
A line with an integrated S/N=7.3 with a well-fitted Gaussian profile (see Fig. \ref{lines_all}) has been 
detected in this spectrum. The detection of just one line at $\nu=88.051$ GHz gives a tentative redshift for
this galaxy and therefore we have to associate this line with the most probable 
transition based on other characteristics of the galaxy. If we associate this detection
with  CO(4--3)  at $z=4.236$, we expect to detect the CO(5--4)
transition at $\nu\sim110.1$ GHz still within our bandpass. The spectrum, however, does not show any line
feature around this frequency and, therefore, this solution is discarded. Other possible
solutions are CO(3--2) or CO(2--1), which
correspond to redshifts of $z=2.927$ and $z=1.618$, respectively. As we expect that the 
probability for  these galaxies to  lie around $z=2-3$  is higher (e.g. \citealt{2005ApJ...622..772C}; 
\citealt{2012MNRAS.420..957Y}; \citealt{2014ApJ...788..125S}; \citealt{2014MNRAS.443.2384Z}), we have  associated our 
line candidate with the CO(3--2) at $z=2.927$, although we cannot rule out the alternative lower redshift solution.

\subsubsection{HLS J102225.9+500536}
For this source we have no detections at S/N$\ge3.5$, either in the raw or in
the binned spectrum. One possible explanation is that
the galaxy lies in the `redshift desert' of the RSR.
However, as we can see in Table \ref{fotometria}, this galaxy has the lowest
flux density at 1.1 mm, which is less than half of those of the rest of the sample. This would suggest
that we need to integrate more in this fainter galaxy to detect a CO emission line. 
Furthermore the ratio $S_{850\mu m}/S_{1.1 {\rm mm}}$ is the largest of the sample,
which could be due to a higher temperature, and make more difficult the  detection of molecular lines
than in the other sources. The upper limit for a line could be calculated using the equation 
$I_{\rm CO}=3\sigma(\delta v\Delta v_{\rm FWHM})^{1/2}$, where $\sigma$ is the rms noise, $\delta v$ the 
velocity resolution, and $\Delta v_{\rm FWHM}$ the line width (e.g. \citealt{2005MNRAS.359.1165G}). 
We adopted a line width of 500 km s$^{-1}$ similar to the lines detected in the remaining spectra 
(see Table \ref{espectros}), and a $\sigma$ derived from the standard deviation of the whole spectrum.
Using this equation the upper limit for a line is 1.2 Jy km s$^{-1}$. 

\subsubsection{HLS J173140.8+225040}\label{MACSJ1731}
This is  another case of single-line detection at $\nu=93.163$ GHz with a line integrated S/N=5.7. In this case
we can associate this transition with CO(4--3) at $z=3.949$ or CO(3--2) at $z=2.712$ having rejected the CO(2--1) 
solution at $z=1.475$ due to the low probability of lensing at this lower redshift. 
Independently, earlier observations made by the HLS team had already identified the redshift of 
this galaxy through a CO redshift search with the IRAM 30-m/EMIR, yielding $z=2.712$ through the detections
of CO(5--4) and CO(3--2) (Dessauges-Zavadsky et al., in preparation), and hence our single line detection unambiguously
corresponds to CO(3--2). 

The line-shape shows some departure from a single Gaussian profile. However, due to the low
pixel-to-pixel S/N, we cannot be certain of the non-Gaussianity, and hence, we decided to perform 
a single Gaussian fit.  If we were to fit two Gaussian profiles, the resulting FWHM for 
both components are $\sim220$ km s$^{-1}$ centred at 93.130 and 93.223 GHz, with a total flux line 
similar to the one recovered with the one fit procedure.

\subsubsection{HLS J204314.2--214439}\label{MACSJ2043}

In this case,  we have detected four lines with S/N$\ge3.5$ in the spectrum at 
$\nu=75.827$, 81.178, 101.459, and 108.428 GHz. They are not compatible
with a single redshift solution, which indicates that we have at least two sources inside  
our RSR beam. 

\begin{figure}
\includegraphics[width=90mm]{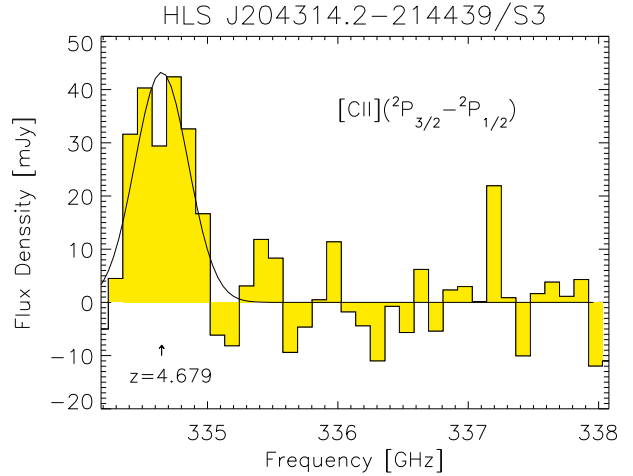}
\caption{[CII] detection of HLS J204314.2-214439/S3 using SMA observations. The redshift derived from this transition,
$z=4.679$, is in good agreement with the redshift derived from the CO LMT-detected lines. This spectrum has 
been obtained after combining the signal from the three lensed images of this galaxy. 
The FWHM and the redshift derived from this detection is in good agreement with the values
derived from the CO emission lines detected with the LMT.}
\label{SMA_CII}
\end{figure}

As a first step to associate these detections with  CO transitions, 
we find that the 81.2 and 101.5 GHz lines agree very well with CO(4--3) and CO(5--4)
at $z=4.680$, however there is no explanation  for the 76 and 108 GHz lines at
this redshift. The $z=4.680$ solution has been confirmed after we reanalysed  
SMA observations, where we found the [CII]
transition at  $\nu=334.648$ GHz (see Fig. \ref{SMA_CII}). 

The Gaussian fitting to the spectral line gives a FWHM of 
$432\pm68$ km s$^{-1}$, which is in good agreement with the line widths of the CO emission lines 
(see Table \ref{espectros}), and a redshift of $z=4.679$ is also in agreement with our redshift 
derived from the CO detections. The integrated flux of the line is $19.1\pm3.6$ Jy km s$^{-1}$.
This source (hereafter HLS J204314.2--214439/S3) is the 
highest redshift galaxy detected by the LMT at this time.

Considering next the 75.827 GHz detection, we reject the association with CO(3--2) at $z=3.56$ since the 
CO(4--3) line should appear at 101.106 GHz. The closest detected
line however is at 101.461 GHz, more than 1000 km/s away, and furthermore we know that this line corresponds to the 
CO(5--4) transition at $z=4.680$. Hence we discard this $z=3.56$ redshift solution. 
Other evidence that the 76 GHz line arises from a different source is the brightness of the line and the asymmetric 
line shape seen in Figs \ref{imagenes} and \ref{lines_all}, whilst the 81 and 101 GHz lines are more 
similar in line width and shape.
We therefore identify the 76 GHz line as  CO(2--1)  at $z=2.040$ (hereafter HLS J204314.2--214439/S1). 
Moreover, the HLS team  independently obtained the same redshift through the detection of 
CO(4--3) and CO(3--2) with IRAM30m/EMIR (Walth et al., in preparation)  and H$\alpha$ with 
Very Large Telescope (VLT)/Spectrograph for Integral Field Observations in the Near Infrared 
(SINFONI, Nakajima et al., in preparation). 

The  remaining 108.428 GHz line does not seem to correspond to any transition at the redshifts of the above galaxies. 
By inspecting the available VLT/SINFONI data of RXCJ2043, Nakajima et al. (in preparation)
found, serendipitously, a strong [OIII] emitter falling within the SCUBA-2, AzTEC, and RSR beams. The simultaneous
detection of the [OIII] doublet ($\lambda = 4959/5007$ \AA) and the H$\beta$ line
confirms an optical redshift of $z=3.25$. This redshift is consistent with our 108.4 GHz line if we
assume  it is due to the CO(4--3) transition. 
At this redshift, we should expect to also detect CO(3--2) at $\sim81.3$ GHz, but this is very close to our previously 
identified line CO(4--3) at $z=4.68$ (HLS J204314.2--214439/S3). In order to test whether there is any evidence of CO(3--2) at 
$z\approx3.25$, we attempt a double Gaussian fit with free central frequencies, widths and amplitudes. The best-fitting 
Gaussian profiles allowed us to recover the CO(3--2) transition at  a redshift consistent with that as predicted by the CO(4--3)
transition at $z=3.252$ and with  a similar FWHM (see Fig. \ref{lines_all} and Table \ref{espectros}).  We conclude that these
two CO lines are most likely associated with a third galaxy at $z=3.252$ (hereafter HLS J204314.2--214439/S2).
In order to estimate the errors for the fits to the CO(3--2) transition we implement a bootstraping method on the noise and 
repeat the Gaussian fit 1000 times. The mean and standard deviation of the fits are reported in Table \ref{espectros}.

The detections of three components at widely different redshifts ($z=2.04$, 3.25, and 4.68) for this blended source 
confirms the results of theoretical works that predict that some of the blended galaxies
are physically unassociated with typical redshift separations $\Delta z\sim0.9-1.5$ (\citealt{2013MNRAS.434.2572H};
\citealt{2015MNRAS.446.1784C}; \citealt{2015MNRAS.446.2291M}).

\subsubsection{Fainter emission lines}\label{CI(1-0)}

Because of the wide bandwidth of the RSR we can search for additional weaker line transitions of different species
that we know to fall within our bandpass, given the redshifts of the galaxies, but which are below
the higher detection threshold for blind searches. In this way,
for HLS J204314.2--214439/S3 the atomic carbon [CI]($^3P_1 \to ^3P_0$) transition at $z=4.680$ is expected at 86.6 GHz, inside
the RSR spectral coverage, where we indeed have evidence for a detection with an integrated line S/N$\approx3$ (see Fig. \ref{lines_all}).
This line has a line width of $\sim 420$ km s$^{-1}$, consistent with the line widths of the 
CO transitions, and an integrated flux of $\sim 0.9 \pm 0.4$ Jy km s$^{-1}$. However, since this detection is 
just at a $3\sigma$ level, measurements derived from this line have been handled, formally, as upper limits.

\subsection{RSR analysis: gas properties}\label{gas_prop}

Having identified the most probable line transitions and redshift of each source, 
we calculate $^{12}$CO luminosities of the galaxies using 
the standard relation given by \citet{2005ARA&A..43..677S}:
\begin{equation}
L'_{\rm CO}=3.25\times 10^7 S_{\rm CO}\Delta V\mbox{ }\nu_{\rm obs}^{-2}\mbox{ }D^2_L\mbox{ }(1+z)^{-3},
\end{equation}
where $L'_{\rm CO}$ is the line luminosity in K km s$^{-1}$ pc$^2$,  $S_{\rm CO}\Delta V$ is the 
velocity-integrated line flux in Jy km s$^{-1}$, $\nu_{\rm obs}$ is the observed central frequency 
of the line in GHz, and $D_L$ is the luminosity distance in Mpc. We 
estimate $S_{\rm CO}\Delta V$ as the integral of the Gaussian fits to the lines, using Monte Carlo 
simulations that take into account the errors in the Gaussian parameters (i.e. peak 
flux density and line width) to estimate the errors.

We continue with an estimation of the H$_2$ mass from the measured $L'_{\rm CO}$ which 
requires two steps. First, luminosities 
originating from higher transitions (J$_{\rm up}\ge2$) must be transformed to an equivalent 
$^{12}$CO(1--0) luminosity using a brightness ratio based on an excitation model. Second, 
once the $L'_{\rm CO(1-0)}$ has been estimated, the H$_2$ mass can be derived by the following 
relationship:
\begin{equation}
M({\rm H_2})=\alpha L'_{\rm CO(1-0)},
\end{equation}
where $\alpha$ is a conversion factor in units of M$_{\sun}$(K km s$^{-1}$ pc$^2$)$^{-1}$.

To transform our $L'_{\rm CO(J_{\rm up}\ge2)}$ to $L'_{\rm CO(1-0)}$ we have used the brightness ratios reported by 
\citet{2013ARA&A..51..105C} for SMGs based on all available literature at the time,
and a value of $\alpha=0.8$ M$_{\sun}$(K km s$^{-1}$ pc$^2$)$^{-1}$ (\citealt{1998ApJ...507..615D}) for the mass 
transformation.  As discussed by \citet{2012MNRAS.424.2429S},  differential magnification could affect the line
ratios of the CO ladder, however, in cluster lensing it is rare to find  strong gradients in the magnification 
on subarcsecond-scales and  therefore the line ratios are usually unaffected (e.g. \citealt{2011MNRAS.410.1687D}).  Furthermore, the higher-J CO emission 
has been shown to be more compact than that of lower-J transitions (e.g. \citealt{2011MNRAS.412.1913I}), which decreases
even further the probability of differential magnification.

The resulting $^{12}$CO luminosities and  H$_2$ masses are reported in Table \ref{espectros2}.
The  mean H$_2$ mass of the sample is $(2.0 \pm 0.2)\times10^{11} M_{\sun}/\mu$, where
$\mu$ is the lens amplification. We caution that these results adopt
average brightness ratios to convert to $L'_{\rm CO(1-0)}$ which introduce extra uncertainties, as well as 
our adopted value of $\alpha$ which is a factor of $\sim4$ lower than the value used in other 
works (e.g. \citealt{2010Natur.463..781T}), and could vary with metallicity and gas properties 
(\citealt{2012MNRAS.421.3127N}).

We also estimated the  luminosity of the $3\sigma$ detection of [CI]($^3P_1 \to ^3P_0$) in HLS J204314.2--214439/S3, finding
a luminosity of $\mu L'_{\rm CI(1-0)}=5.2\pm2.6\times10^{10}$ K km s$^{-1}$ pc$^2$.
Comparing this value with the $^{12}$CO transitions detected for this galaxy, we have  line ratios of 
$L'_{\rm CI(1-0)}/L'_{\rm CO(4-3)}=0.21\pm0.11$ and $L'_{\rm CI(1-0)}/L'_{\rm CO(5-4)}=0.56\pm0.29$. These values are 
consistent with other measurements of SMGs and quasars and do not differ significantly from what is found in 
low-redshift systems, as discussed by  \citet{2011ApJ...730...18W} (see also 
\citealt{2013MNRAS.435.1493A}; \citealt{2014MNRAS.444.1301P}).

\begin{table*}
\caption{Physical properties derived from the $^{12}$CO lines detected in the RSR spectra and
the SCUBA-2+AzTEC photometry. Column 1: name of the source; column 2: observed CO luminosity; column 3:
CO(1--0) luminosity converted using the factors reported by \citet{2013ARA&A..51..105C}; column 4: 
$\rm H_2$ mass assuming $\alpha=0.8$ M$_{\sun}$ (K km s$^{-1}$ pc$^2$)$^{-1}$ (\citealt{1998ApJ...507..615D}); 
column 5: dust masses estimated from the flux densities at 850 $\mu$m and 1.1 mm; 
column 6: total FIR luminosity; column 7: amplification factor.}
\begin{tabular}{lcccccc}
\hline
Source& $\mu L'_{\rm CO(J_{\rm up})}$ & $\mu L'_{\rm CO(1-0)}$  &$\mu$M$(\rm H_2)$& $\mu M_{\rm d}$ &$\mu L_{\rm FIR}$&$\mu$\\
 & ($\times10^{10}$ K km s$^{-1}$ pc$^2$) &($\times10^{10}$ K km s$^{-1}$ pc$^2$)& ($\times10^{10}$ M$_{\sun}$) & ($\times10^{8}$ M$_{\sun}$)&($\times10^{12}$ L$_{\sun}$)&\\
\hline
HLS J045518.0+070103      & $20.4\pm3.6$ &$31.0\pm5.5$  &$24.8\pm4.4$   &$25\pm2.0$     &$63^{+5}_{-7}$   &$\sim 4$  \\
HLS J173140.8+225040      & $8.9 \pm3.0$ &$13.5 \pm4.5$ &$10.8\pm3.6$   &$36\pm2.0$     &$54^{+5}_{-4}$   &$\sim 2$  \\
HLS J204314.2--214439/S1   & $32.4\pm5.0$ &$38.1\pm5.8$  & $30.5\pm4.7$  &$\sim15^a$     &$\sim25^a$       &$\sim 5$\\
HLS J204314.2--214439/S2   & $2.6\pm1.6$  &$3.9\pm2.3$   & $3.2\pm1.9$   &$\sim13^a$     &$\sim33^a$       &$\sim 3^b$\\
                          & $3.8\pm1.2$  &$8.3\pm2.5$   & $6.7\pm2.0$   &$\sim13^a$     &$\sim33^a$       &$\sim 3^b$\\
HLS J204314.2--214439/S3   & $24.9\pm3.2$ &$54.2\pm6.9$  &$43.3\pm5.6$   &$\sim11^a$     &$\sim44^a$       &$\sim 3^b$ \\
                          & $9.3\pm1.5$  &$23.7\pm3.9$  &$19.0\pm3.1$   &$\sim11^a$     &$\sim44^a$       &$\sim 3^b$  \\
\hline
\multicolumn{7}{l}{$^a$ The uncertainties in these values are large since the real contribution from each galaxy to the total detected (blended) flux}\\
\multicolumn{7}{l}{is unknown. We assume equal contribution to the (sub-)mm fluxes by all components. $^b$Average  of the two transitions.}\\
\label{espectros2}
\end{tabular}
\end{table*}

\begin{figure}
\includegraphics[width=90mm]{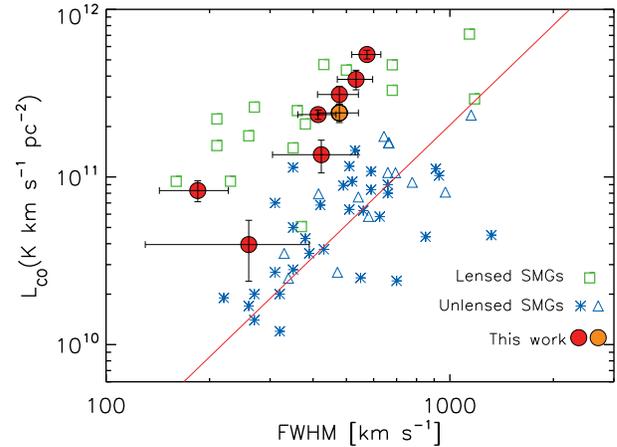}
\caption{$L'_{\rm CO(1-0)}$ versus line FWHM for lensed and unlensed SMGs. Red circles represent
the measurements from this work and light red is used when we adopt the alternative redshift solution for
HLS J045518.0+070103.  Estimates derived from each detected transition are plotted individually. 
Blue triangles are unlensed or lensing-corrected SMGs with CO(1--0)
measurements (\citealt{2010ApJ...714.1407C}; \citealt{2010ApJ...723.1139H}; \citealt{2011MNRAS.412.1913I}; 
\citealt{2011ApJ...739L..31R}; \citealt{2013Natur.498..338F}) and blue asterisks are SMGs with
higher-J CO line measurements converted to CO(1--0) from \citet{2013MNRAS.429.3047B}. The red line shows
the best-fitting relation for unlensed SMGs $L'_{\rm CO}=10^{5.4}{\rm FWHM}^{2}$ derived by 
\citet{2013MNRAS.429.3047B}. The green squares are GBT and VLA CO(1--0) detections of lensed SMGs 
(\citealt{2012ApJ...752..152H}; \citealt{2014ApJ...783...59R}).}
\label{LCO_FWHM}
\end{figure}

As described by \citet{2012ApJ...752..152H}, the line luminosity-line width relation, 
$L'_{\rm CO(1-0)}=a(\Delta v_{\rm FWHM})^b$ (see Fig. \ref{LCO_FWHM}),
could be used as an estimator of the lens magnification, $\mu$, if we assume that the magnification only 
modifies the observed  line luminosity (and not the line width). This means that $L'_{\rm apparent}=\mu L'$ or 
\begin{equation}
\mu=\frac{L'_{\rm apparent}}{L'}=\frac{L'_{\rm apparent}}{a(\Delta v_{\rm FWHM})^b}.
\end{equation}

Our measurements lie within the same region occupied by other lensed SMGs in Fig. \ref{LCO_FWHM},  which  provides circumstantial 
evidence that amplification due to lensing is present. Table \ref{espectros2} lists the lens magnifications derived from this 
equation, adopting the line luminosity--line width relation for unlensed SMGs derived by  \citet{2013MNRAS.429.3047B}
(see also \citealt{2012ApJ...752..152H}).
We measure modest amplification, $\mu\approx2-5$, for the sample.

\subsection{SCUBA-2 + AzTEC analysis: dust properties}

The dust continuum detections allow us to estimate the dust temperature  and IR luminosity
once a spectral energy distribution (SED) is fitted to the data. Our determinations of spectroscopic redshifts 
(see Section \ref{line_ident}) break the temperature--redshift degeneracy 
(e. g. \citealt{2002PhR...369..111B}), albeit there is still a degeneracy between temperature and spectral
index (e.g. \citealt{1993MNRAS.263..607H}; \citealt{2014arXiv1402.1456C}). 

Assuming the dust  is isothermal, the dust mass, $M_{\rm d}$, is estimated from 
\begin{equation}
M_{\rm d}=\frac{S_\nu D_L^2}{(1+z)\kappa_\nu B(\nu,T_{\rm d})},
\end{equation}
where $S_\nu$ is the  flux density at frequency $\nu$, $\kappa_\nu$ is the dust
mass absorption coefficient at $\nu$, $T_{\rm d}$ is the dust temperature,
and $B(\nu_,T_d)$ is the Planck function at temperature $T_{\rm d}$. The dust mass 
absorption follows the same power law as the optical depth, $\kappa\propto \nu^\beta$ 
with a normalization of $\kappa_{\rm d}(850\mu{\rm m})=0.15$ m$^2$ kg$^{-1}$ (\citealt{2003Natur.424..285D}).

We fit modified blackbody functions fixed at the derived redshift of each galaxy,
to the continuum photometry data in Table \ref{fotometria} along with the SPIRE photometry, in order to 
estimate the dust temperature. For the HLS J204314.2--214439 blended galaxies we adopt a dust temperature of 
$T=37\pm8$ K as that measured in lensed galaxies (\citealt{2013ApJ...767...88W};
see also \citealt{2012ApJ...752..152H}) since the contribution from each galaxy to the total blended flux 
is unknown. This value is consistent with the estimated temperatures of the other galaxies in our sample (e.g. $38\pm3$
and $35\pm3$ K, for HLS J045518.0+070103 and HLS J173140.8+225040, respectively).

Using this temperatures  and a fixed emissivity index of  $\beta=1.7$ (empirical fits to 
the observed long wavelength SEDs suggest $\beta=1.5-2$; e.g. \citet{2001MNRAS.327..697D}; 
\citealt{2009MNRAS.398.1793C}; \citealt{2012A&A...539A.155M}), we have derived the dust mass 
for each galaxy from the flux density at both 850 $\mu$m 
and 1.1 mm, and extrapolating $\kappa$ to the observed rest-frame for each galaxy. These values 
are reported in Table \ref{espectros2} and have a  mean  dust mass of 
($2.0\pm0.2)\times10^9$ M$_{\sun}/\mu$. If we extrapolate the observed spectral index to use 
$\kappa_d(850\mu{\rm m})$ directly, the mean dust mass increases by a factor of $\sim 2$. These calculations do not include  
uncertainties in $\kappa$, which could be at least a factor of 3 (\citealt{2003Natur.424..285D}). 
For HLS J204314.2--214439/S1, HLS J204314.2--214439/S2, and HLS J204314.2--214439/S3  we have assumed they each contribute a third 
of the total (blended) flux density, as a first approximation. A full multi-wavelength analysis of this triple system,
including interferometric observations, will be presented by Walth et al. (in preparation).

Finally, the estimated FIR luminosites ($L_{\rm FIR}$; $42.5-122.5 \mu$m), which arise from the SED fitting,
are reported in Table \ref{espectros2} and discuss below.

\section{DISCUSSION}

\subsection{The L$'_{CO}$-L$_{FIR}$ relation}

The SFE, with which the molecular gas is being
turned into stars, can  be inferred from our observations. This quantity is often 
estimated by the ratio SFR/$M(\rm {H}_2$) -- the inverse of the gas 
depletion time -- however, a more straightforward estimate of 
the SFE is the continuum-to-line luminosity ratio, 
$L_{\rm FIR}/L'_{\rm CO(1-0)}$, because it does not depend on
the CO--$H_2$ conversion factor, $\alpha$. 
This describes, in observational terms, the relationship
between the luminosity due to star formation and the gas content. 

\begin{figure}
\includegraphics[width=90mm]{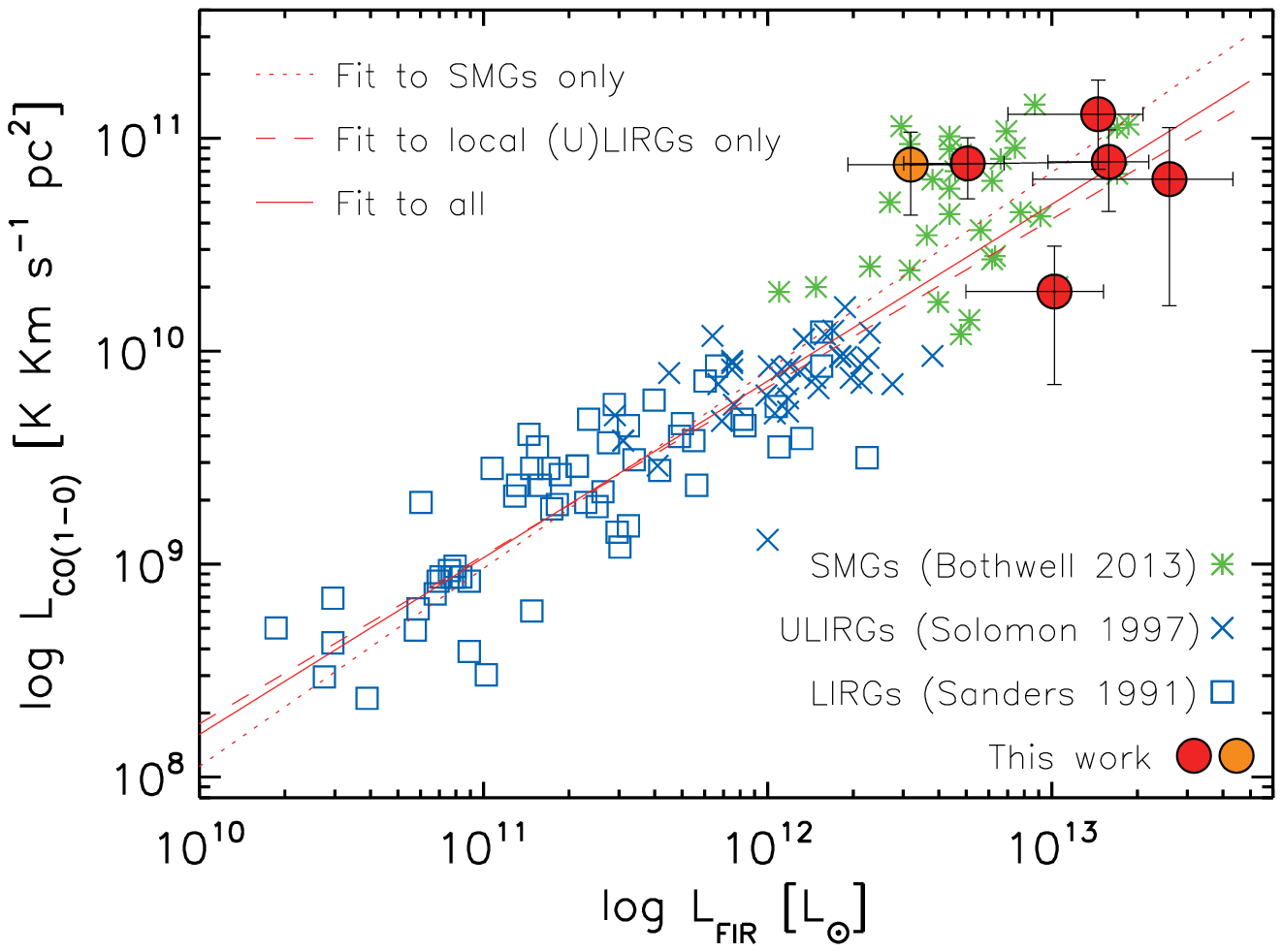}
\caption{Lens-corrected CO luminosity versus FIR luminosity 
($L'_{\rm CO}-L_{\rm FIR}$) for our sample, in red circles, and in light red
if we adopt the alternative CO solution for HLS J045518.0+070103. For galaxies with more than
one detected transition we have  plotted the mean $L'_{\rm CO}$. For comparison,
other SMGs are plotted (\citealt{2013MNRAS.429.3047B}) along 
with local LIRGs (\citealt{1991ApJ...370..158S}) and ULIRGs
(\citealt{1997ApJ...478..144S}). Best-fitting slopes derived by 
\citet{2013MNRAS.429.3047B}  to  SMGs,  local (U)LIRGs,  and 
all three combined samples are overplotted. They have slopes of $0.93\pm0.14$, 
$0.79\pm0.08$, and $0.83\pm0.09$, respectively.}
\label{LCO_LFIR}
\end{figure}

In Fig. \ref{LCO_LFIR} we  plot our galaxies with CO 
detections, after correcting for magnification, on to the 
$L'_{CO}-L_{FIR}$ plane, including observations of local (ultra-) luminous infrared 
galaxies (U)LIRGs (\citealt{1991ApJ...370..158S}; \citealt{1997ApJ...478..144S}) and other
SMGs (\citealt{2013MNRAS.429.3047B}). We also show the three power-law
fits derived by \citet{2013MNRAS.429.3047B} to the local (U)LIRGs alone
(with a slope of $0.79\pm0.08$), the SMGs alone ($0.93\pm0.14$), and the
combined sample ($0.83\pm0.09$). Our results 
are consistent with other SMG observations, but they are also consistent with the fit to 
the combined sample and the (U)LIRGs alone. Furthermore, as discussed by 
\citeauthor{2013MNRAS.429.3047B}, our analysis requires extrapolating 
from high-$J_{\rm up}$ $^{12}$CO transitions to $^{12}$CO(1--0), which introduces
extra uncertainties without brightness temperature ratio measurements for 
individual sources. Although we have 
used the estimated amplification factors, the ratio between CO and FIR luminosity
is independent of magnification, assuming no differential amplification, which is
usually the case in the cluster lensing.

\subsection{The gas-to-dust ratio}

Constraining the molecular gas properties of a large sample of  non-gravitationally amplified SMGs
with spectroscopic observations requires a significant  investment of time at present
(\citealt{2013MNRAS.429.3047B}). An alternative approach to estimate the molecular gas mass is to 
use the continuum emission of the dust in order to estimate the dust mass, 
and then use an appropriate gas-to-dust ratio, $\delta_{\rm GDR}$, to derive the total mass of the 
molecular ISM. However, this gas-to-dust ratio is only well constrained in Milky 
Way molecular clouds and  local galaxies (e.g. \citealt{2007ApJ...663..866D}),
and only in a few SMGs (e.g. \citealt{2013arXiv1310.6362S}). 

Combining our estimation of H$_2$ gas mass  with the dust mass resulting from the SCUBA-2
and AzTEC flux densities, we estimate the gas-to-dust ratio for these high-redshift 
galaxies, which is independent of the lensing amplification, assuming that both gas and dust are 
amplified by the same factor, which should be applicable to these cluster-lensed sources. 
Using our independent estimation of these quantities we have found a weighted  mean 
gas-to-dust  ratio of  $\delta_{GDR}\approx55$ for the two galaxies without direct evidence of
blending (HLS J045518.0+070103 and HLS J173140.8+225040). If we also consider the blended galaxies,
assuming they all contribute equally to the total integrated flux with intrinsic flux density errors of
50 per cent, we find a weighted mean $\delta_{\rm GDR}\approx75$. The scatter in these measurements is 
however high ($\delta_{\rm GDR}=20-390$), especially for the blended sources. Our estimated mean values 
are consistent with those of non lensed SMGs, $\delta_{\rm GDR}=90\pm25$ (\citealt{2013arXiv1310.6362S}), 
and also with the lensed $z=6.3$ HFLS3 galaxy, $\delta_{\rm GDR}\sim80$ (\citealt{2013Natur.496..329R}). 
 These ratios are smaller than local values, as also discussed  in the literature  
(e.g. \citealt{2006ApJ...650..592K}; \citealt{2010A&A...518L.154S}; \citealt{2014arXiv1408.0816D}), 
which supports the `dust richness' interpretation of SMGs. For reference, the Milky Way
has a gas-to-dust ratio of $\sim135-185$ (Draine et al. 2007), and local LIRGs  $120\pm28$ 
(\citealt{2008ApJS..178..189W}). We emphasize, however, that the scatter in our measurements, 
especially for the blended sources, is high, and some of our estimated  individual ratios for the
blends overlap with the local values.

\section{Summary}

We present AzTEC 1.1 mm continuum observations and RSR spectra taken with the LMT, as part of the LMT 
Early Science campaign, of a sample of four cluster-lensed SMGs. 

We surprisingly find that one of the four targets studied is a blend of three
galaxies at different redshifts. This confirms that the multiplicity that has been found by 
high-resolution interferometric observations in normal SMGs (\citealt{2011ApJ...726L..18W}; \citealt{2012A&A...548A...4S}; \citealt{2013ApJ...768...91H}) 
is also present in lensed galaxies,  although our search focuses on spectroscopically distinct
components. This also confirms the theoretical predictions of physically unassociated components 
in blended galaxies (e.g. \citealt{2013MNRAS.434.2572H}; \citealt{2015MNRAS.446.1784C}; \citealt{2015MNRAS.446.2291M}).
The sample, though,  is too small to derive firm conclusions on the statistics of 
spectroscopic multiplicity.

Of the five reported redshifts, just one was obtained  with the robust blind detections 
of two lines within a single RSR spectrum. Three other redshifts were confirmed thanks to the
detection of different transitions with other telescopes, and the remaining one is a tentative solution.

The estimated mean H$_2$ gas mass estimated from the CO  lines detected is $(2.0 \pm 0.2)\times10^{11} M_{\sun}/\mu$
and the mean dust mass estimated from the continuum data is $(2.0 \pm0.2)\times10^{9} M_{\sun}/\mu$, 
where $\mu$ is the lensing factor. These results are self-consistent using either the 850 $\mu$m or 1.1
mm data. Using the line width--luminosity relation for SMGs from \citet{2013MNRAS.429.3047B} we 
estimate $\mu\sim 2-5$.

We infer from our independent estimations of gas and dust masses  a weighted mean gas-to-dust ratio 
$\delta_{\rm GDR}\approx55-75$, which, given the little likelihood of differential amplification 
at mid-J transitions, should be independent of the amplification factor. These values
are consistent with other measurements of SMGs and lower than the estimated ratios for local galaxies.

Finally, we estimate the FIR luminosity using our photometric data and with this the SFE of our 
galaxies through the $L'_{CO}-L_{FIR}$ relationship. Our bright galaxies 
follow the correlation found for local (U)LIRGs and other SMGs, with a close to linear slope. 
This suggests that, at the largest luminosities sampled by our systems, the SFE is maintained arguing 
for a universal value from the lowest to the highest redshifts and from the smallest to the
largest mass reservoirs of molecular gas.

The successful blind redshift search in these bright SMGs, performed during this LMT Early Science 
operation period with the current 32-m diameter illuminated surface, highlights the competitiveness of this new 
facility. The sensitivity that will be achieved with the expanded primary surface of the full 50-m diameter LMT will place
this telescope among the most powerful facilities to perform spectroscopic CO surveys (see 
\citealt{2011ApJ...726L..22F} for a comparison of different telescopes), allowing us to
estimate redshifts for significant samples of highly obscured dusty star-forming galaxies 
which are not measurable with even the largest optical and near-IR telescopes.

\section*{Acknowledgments} 
We would like to thank an anonymous referee for a constructive and helpful report.

This work would not have been possible without the longterm financial support from the Mexican Science and
Technology Funding Agency, Consejo Nacional de Ciencia y Tecnolog\'ia (CONACYT) during the construction
and early operational phase of the Large Millimeter Telescope {\it Alfonso Serrano}, as well as support from the the
US National Science Foundation via the University Radio Observatory program, the Instituto Nacional de 
Astrof\'isica, \'Optica y Electr\'onica (INAOE) and the University of Massachusetts (UMASS).

This work has been mainly supported by Mexican CONACyT research grants
CB-2011-01-167291 and CB-2009-133260, JAZ is also supported by a CONACyT studentship. JEG acknowledges the Royal Society.
IS acknowledges support from STFC (ST/L00075X/1), the ERC Advanced Investigator programme DUSTYGAL
(321334)  and a Royal Society/Wolfson Merit Award. 
AMS acknowledges an STFC advanced fellowship through grant ST/H005234/1.

The James Clerk Maxwell Telescope is operated by the Joint Astronomy Centre on behalf of the Science
and Technology Facilities Council of the United Kingdom, the National Research Council of Canada, and
(until 2013 March 31) the Netherlands Organisation for Science Research. Additional funds for the 
construction of SCUBA-2 were provided by the Canada Foundation for Innovation.

The Submillimeter Array is a joint project between the Smithsonian Astrophysical Observatory and the
Academia Sinica Institute of Astronomy and Astrophysics and is funded by the Smithsonian Institution
and the Academia Sinica.

\bsp

\label{lastpage}

\end{document}